\documentclass[12pt]{iopart}

\usepackage{graphicx}

\expandafter\let\csname equation*\endcsname\relax
 \expandafter\let\csname endequation*\endcsname\relax
\usepackage{amsmath}
\usepackage{amsthm}
\usepackage{soul}
\usepackage{color}
\usepackage[usenames,dvipsnames]{xcolor}

\sethlcolor{yellow}
\usepackage{iopams}

\begin{document}

\title[Formation of structure in growing networks]{On the formation of structure in growing networks}

\author{P Moriano and J Finke}
\address{Department of Electrical Engineering and Computer Science, Pontificia Universidad Javeriana, Santiago de Cali, Colombia}
\eads{\mailto{moriano@ieee.org}, \mailto{finke@ieee.org}}
\begin{abstract}
Based on the formation of triad junctions, the proposed mechanism generates networks that exhibit extended rather than single power law behavior. Triad formation guarantees strong neighborhood clustering and community-level characteristics as the network size grows to infinity. The asymptotic behavior is of interest in the study of directed networks in which $(i)$ the formation of links cannot be described according to the principle of preferential attachment; $(ii)$ the in-degree distribution fits a power law for nodes with a high degree and an exponential form otherwise; $(iii)$ clustering properties emerge at multiple scales and depend on both the number of links that newly added nodes establish and the probability of forming triads; and $(iv)$ groups of nodes form modules that feature less links to the rest of the nodes.
\end{abstract}



\maketitle

\section{Introduction}

Networks are systems composed of well-defined elements (nodes) that display collective behaviors at multiple levels of analysis. Large networks arise by the gradual addition of elements which attach to an existing and often evolving network component. With our modern access to data, the application of network techniques offers a wide set of mathematical tools to visualize data at the level of the data elements and the interaction between them. These tools allow us to characterize higher-level properties of the structure of a system and to identify different types of patterns in the relationships among elements.

The development of models that describe the evolution of networks has been driven by the need to analyze large amounts of relational data across a wide range of fields. Well-known examples include the study of relationships we see in scientific collaborations {\cite{Newman:2001fk}}, export goods {\cite{Hidalgo:2007uq}}, traffic {\cite{barrat:2004uq}}, social ties {\cite{Kossinets:2006fk}}, stocks {\cite{Onnela:2003uq}}, and patent citations {\cite{Hall:2001uq}}-{\nocite{Valverde:2007uq}}{\cite{Moriano:2011fk}}. Trying to address the question of how particular topologies arise as networks grow, a large body of work has been devoted to understand the emergence of three properties: the distribution of links per node (degree distribution), the proportion of links grouped into local neighborhoods (clustering or transitivity) {\cite{Albert:2002zl}}, {\cite{Newman:2003fv}}, and the division of the set of nodes into modules (communities) with tight interconnections within and sparser links across them {\cite{Newman:2004fk}}.

In extended power law networks, the probability $p_k$ that a node with a low degree of connectivity (below some threshold $\varepsilon$) connects to $k$ other nodes fits an exponential form $e^{-\lambda k}$ for some positive constant ${\lambda}$. For nodes with a high degree, the probability $p_k$ is proportional to the power law function $k^{-\alpha}$ for some positive constant $\alpha$. Because the tail of the probability distribution of the degree of the nodes has no exponential bound, the patterns of interaction in power law networks differ in orders of magnitude, with a few nodes being highly connected. Mechanisms leading to power law networks have been overviewed in {\cite{Mitzenmacher:2004fk}}. A particular class of mechanisms in which nodes with a high degree have a greater probability of acquiring new links (attributed to the principle of preferential attachment) has been proposed to explain the scaling behavior in empirical data {\cite{Barabasi:1999fr}}, {\cite{Reed:2002fk}}.

In clustered networks, the probability of finding transitive triplets is higher than the outcome expected through random chance. If a node connects to two other nodes, clustering captures the probability that these two nodes are connected, too. In a network with high clustering, nodes do not interact homogeneously with other nodes, but tend to influence each other locally (i.e., they form strong neighborhood clusters {\cite{Louch:2000fk}}). Common measures of clustering are based on $(i)$ the total number of transitive triplets relative to the total number of possible triplets in the network, represented by a global clustering coefficient $C$ {\cite{Newman:2003fv}}; or $(ii)$ the fraction of triplets connecting the neighboring nodes of node $i$ over the total number of possible triplets, represented by a local clustering coefficient $C_i$ {\cite{Watts:1998rr}}. Real-world networks show clustering coefficients that are generally independent of the size of the network and scale with the degree of the nodes {\cite{Ravasz:2003fk}}.

In networks with community structure, the division of the set of nodes into modules underlies their dynamic formation. Nodes may group according to particular characteristics (types), reflecting a tendency to establish stronger ties with similar others (e.g., according to interests, occupation, or beliefs) \cite{Guerrero:2012uq}. Under this proposition, the modularity of a network captures the difference between the average fraction of edges within communities and the expected value for a random network ($Q$-modularity) \cite{Newman:2004fk}. A measure of modularity $Q > 0.3$ suggests the existence of a well-defined community structure (often found in social and information networks).

Though preferential attachment offers an explanation for the existence of networks with power law degree distributions, it does not, by itself, explain the formation of strong neighborhood clusters. Clustering coefficients tend to vanish with the continuous addition of new nodes (based on both local and global preferential attachment mechanisms {\cite{Li:2003fk}}). The development of alternative models that can explain strong neighborhood clustering as the natural outcome of the process of growth contributes towards establishing a framework that supports the analysis of the clustering behavior of power law networks.

Based on the principle of preferential attachment, the authors of {\cite{Holme:2002uq}}, {\cite{Zhang:2007uq}} introduce a baseline probability of establishing additional links by a process of triad formation. They generate undirected networks with tunable degree distributions and clustering properties. In {\cite{Zhang:2007uq}} the authors deduce analytical results based on  generic conditions underlying local attachment mechanisms. Unlike {\cite{Holme:2002uq}}, {\cite{Zhang:2007uq}} the work in {\cite{Jackson:2007uq}} explains power law behavior in networks in which the process of establishing links does not necessarily depend on preferential attachment. The attachment of new nodes results according to a uniform random distribution followed by the formation of triad junctions. Like {\cite{Jackson:2007uq}} the formation mechanism in this paper does not instantiate the principle of preferential attachment.

Although the model in {\cite{Jackson:2007uq}} generates extended (rather than single) power laws in the in-degree distribution of strongly clustered networks {\cite{Shao:2006fk}}, {\cite{Moriano:2012uq}}, it does not describe the threshold that marks the transition from an exponential fit to a power law. Here, we deduce analytical expressions for $(i)$ the exponential exponent $\lambda$ that characterizes the behavior of nodes with a low degree; $(ii)$ the threshold $\varepsilon$ above which nodes follow a power law; and $(iii)$ the relationships between the clustering coefficients and the value of $\varepsilon$. The expressions for the degree distribution and the clustering coefficients (all dependent on $\varepsilon$) imply that there exist common factors driving the formation of structure during network growth. Unlike the work in \cite{Jackson:2007uq}, the proposed mechanism rests on an immediate implementation of the principle of triad formation (i.e., triads may be formed after every random attachment, as opposed to forming triads by choosing a node from the union of the set of all neighbors after all random attachments). This difference in the process of triad formation yields expressions for both global and local clustering coefficients which unlike the expressions in \cite{Jackson:2007uq} depend on the threshold $\varepsilon$.

The contribution of the proposed mechanism is threefold. First, it explains scaling behavior in networks with an extended power law in their in-degree distribution (offering a better fit than a single or double power law distribution to describe social and information networks \cite{Valverde:2007uq}, \cite{Csanyi:2004fk}). Second, it accounts for strong neighborhood clustering based on a random triad formation process with a positive stationary mean probability. Clustering properties remain constant as the size of the network grows to infinity. Third, it explores the formation of communities from allowing nodes to establish stronger ties with nodes of the same type (group preference).

The remaining sections are organized as follows. First we introduce a model of the connectivity of a network that grows through the continuous addition of new nodes. Theorem~1 shows that, above a certain threshold $\varepsilon$, the in-degree distribution follows a power law distribution with scaling exponent $\alpha$, and an exponential distribution with exponential exponent $\lambda$, otherwise. We present analytical results for values of $\alpha$, $\lambda$, and $\varepsilon$. The results suggest that the transition from exponential to power law distributions depends on both the scaling exponent (which, in turn, depends on the probability of forming triads) and the number of links that newly added nodes establish. Theorem~2 characterizes the evolution of the global and local clustering coefficients and presents asymptotic expressions for $C$ and $C_i$ (both depend on $\varepsilon$). Second, we characterize the relationship between the formation of triads and the scaling exponent of the network. Simulations also show the effect of group preference and network modularity on $\alpha$, $\lambda$, $C$, and $C_i$. Third, we apply the proposed mechanism to generate realizations that resemble the degree distribution and clustering properties of an empirical network with no directed cycles. In particular, we consider the opinions written by the U.S. Supreme Court and the cases they cite \cite{Fowler:2008fk}. We discuss how the model contributes to the understanding of the semantic evolving topology, and more generally, how it identifies generic conditions that lead to the formation of structure as these types of acyclic directed networks grow. Finally, we draw some conclusions and future research directions.

\section{A network formation model} 

Let the graph $\mathcal{G}_t=(\mathcal{H}_t,\mathcal{A}_t)$ represent the network at time index $t$. The set $\mathcal{A}_t = \{(i,j): i,j \in \mathcal{H}_t\}$ represents the relationships between a finite set of interconnected nodes that belong to $\mathcal{H}_t = \{1, \ldots, N_t\}$. The pair $(i,j)$ indicates that there exists a directed edge between nodes $i$ and $j$, and $q_i(t) = \{j \in \mathcal{H}_t :(j,i) \in \mathcal{A}_t\}$ represents all nodes that link to node $i$ (i.e., its incoming neighbors at time $t$). For any node $i \in \mathcal{H}_t$, let $k_{i}(t) = \vert q_i(t)\vert$ represent the in-degree of node~$i$. 

\subsection{Node attachment}

Every time index $t$ a new node attaches to $m$ different nodes, selected according to a uniformly random distribution over $\mathcal{H}_{t-1}$. Let $n \ge 0$ denote the number of edges established from nodes in $\mathcal{H}_{t-1}$ to the newly added node, according to some mechanism that responds to the attachment. If there is no such response underlying the node attachment process then $n = 0$ (e.g., for a network with no directed cycles).

\subsection{Triad formation}

The requirements for the formation of triad junctions are similar to the conditions introduced in \cite{Holme:2002uq}. When node $j \notin \mathcal{H}_{t-1}$ attaches to some node $j' \in \mathcal{H}_{t-1}$, it may also establish an additional link to one of the outgoing neighbors of node $j'$, selected again according to a uniformly random distribution. If $j \in q_{j'}(t)$ and $j' \in q_{i}(t)$ for some node $i$, node $j$ links to node $i$ with probability $x_{i} (t)$. A multivariate random variable $X_t$ with a positive expected probability $p_t=E[X_t]= f(\sigma_1, \ \cdots, \ \sigma_s) d \sigma_1 \cdots d \sigma_s$ captures the set of possible different probabilities of establishing a link between nodes $j$ and $i$, where $\sigma_1, \ \cdots, \ \sigma_s$ are independent factors that influence the formation of triads. Note that if the set of outgoing neighbors of node $j'$ is a subset of the set of outgoing neighbors of node $j$ then there is no possibility of establishing additional links through triad formation. The process repeats for every edge established by a newly added node ($m$ times) before another node may attach to the network. Let $X=\{ X_t\}$ with stationary mean $p > 0$ be the random process associated to the process of triad formation. \\

\noindent \emph{Assumption~$1$ (on the initial network)}: To ensure that the two-step mechanism (growth-plus-triad-formation) can be properly completed, we require that $(a)$ the network $\mathcal{G}_0$ is weakly connected; and $(b)$ the network $\mathcal{G}_0$ has at least $m$ nodes, each with at least one outgoing neighbor. \\

\noindent Assumption~1$(a)$ is satisfied if replacing all the directed edges with undirected ones produces a connected undirected graph. Assumption~1$(b)$  means that $N_0 \ge m$ and for every node $i\in \mathcal{H}_0$ there exists a node $i'$ such that $i\in q_{i'}(0)$. This last condition is required when $p=1$.

\section{Analysis}

It is of interest that the mechanism guarantees topological properties of both the in-degree distribution and the clustering coefficients of the network. \\

\noindent \emph{Theorem $1$ (in-degree distribution): For all $\mathcal{G}_0$ that satisfy Assumption~$1$, the in-degree distribution $p_k$ of $\mathcal{G}_t$ follows an extended power law as $t \to \infty$. The scaling and exponential exponents are $\alpha=2+ \frac{1}{p}$ and $\lambda=\frac{\alpha}{\alpha-1}$ with threshold~$\varepsilon=\left( \alpha -1  \right) m$.} 
\begin{proof}
We assume that the in-degree of node $i$ is a continuous variable $k_i \in \mathbb{R}, \ k_i \ge 0$. Every time index $t$ a newly added node $j \notin \mathcal{H}_{t-1}$ attaches to $m$ different nodes in $\mathcal{H}_{t-1}$, selected according to a uniform distribution process over the $N_0 + t - 1$ existing nodes. The probability that node $j$ attaches at time $t$ to a node $i \in \mathcal{H}_{t-1}$ is
\begin{equation*}\label{e1}
\frac{m}{N_0 + t-1}
\end{equation*}
The triad formation step that (immediately) follows random attachment adds to the rate of change of node $i$ with in-degree $k_i (t-1)$ by 
\begin{equation*}\label{e2}
\left( \frac{m k_i (t)}{N_0+t-1}\right) \left(\frac{1}{m(1+p)}\right) p
\end{equation*}
The first term $\frac{m k_i (t)}{N_0+t-1}$ is the probability of selecting, during random attachment, an incoming neighbor of node $i$ (i.e., some node $j' \in q_i (t)$). The second term $\frac{1}{m(1+p)}$ is the probability that node $j'$ is an incoming neighbor of node $i$ (i.e., $j' \in q_i (t)$). Furthermore, the probability $p$ is the stationary mean of the random process of forming triads. The multiplication of all 3 terms define the probability of forming a triplet with an edge that contributes to the in-degree of node $i$. Thus, the overall rate of change of $k_i(t)$ is
\begin{equation}\label{e3}
\frac{d k_i (t)}{dt}=\frac{m}{N_0 + t-1}+\frac{p}{1+p}\frac{k_i(t)}{N_0+t-1}
\end{equation}
with boundary condition $k_i (t_i)=n$. The solution to~\eqref{e3} is 
\begin{eqnarray}\label{e5}
k_i (t) &=& \left (n + \left (1 + \frac{1}{p} \right ) m \right )\left( \frac{N_0 + t-1}{N_0 + t_i -1} \right)^{\frac{p}{1+p}} \nonumber \\ 
&&- \left (1 + \frac{1}{p} \right )m
\end{eqnarray}
Using~\eqref{e5}, the analytical expression for the cumulative distribution of the in-degree $P[k_i (t) \le k]$ of node $i$ equals 
\begin{eqnarray}\label{e6}
&&P\left [\left (n + \left (1 + \frac{1}{p} \right ) m \right )\left( \frac{N_0 + t -1}{N_0 + t_i -1} \right)^{\frac{p}{1+p}} \right. \nonumber \\
&&- \left. \left (1 + \frac{1}{p} \right )m \le k \right] \nonumber \\
&=&P \left [t_i \ge \left( \frac{n+ \left (1 + \frac{1}{p} \right )m}{k+  \left (1 + \frac{1}{p} \right ) m} \right)^{1+\frac{1}{p}} (N_0 + t -1) \right.  \nonumber \\ 
&&- \left. (N_0-1) \right  ] \phantom \nonumber
\end{eqnarray}
And as $t \to \infty$
\begin{eqnarray}\label{e55}
P[k_i (t) \le k]=1-\left( \frac{n + \left (1 + \frac{1}{p} \right) m}{k + \left (1 + \frac{1}{p} \right) m} \right)^{1+\frac{1}{p}}
\end{eqnarray}
Finally,
\begin{eqnarray}\label{e7}
p_k=\frac{d P[k_i (t) \le k]}{dk}=a\left(k+\left (1 + \frac{1}{p} \right )m \right)^{-\left(2+\frac{1}{p}\right)} 
\end{eqnarray}
where $a=\left(1+\frac{1}{p}\right) \left (n + \left (1 + \frac{1}{p} \right )m\right)^{1+\frac{1}{p}}$. Note that~\eqref{e7} exhibits an extended power law of the form
\begin{eqnarray*}\label{e12}
p_k\sim (k+\varepsilon)^{- \alpha}
\end{eqnarray*} 
where $\alpha=2 + \frac{1}{p}$ and $\varepsilon =\left (\alpha-1 \right )m$. When $k \gg \varepsilon$,~\eqref{e7} is reduced to a single power law $p_k \sim k^{-\alpha}$. On the other hand, when $k \ll \varepsilon$ we have
\begin{eqnarray*}\label{e13}
\ln p_k \sim - \alpha \ln(k+\varepsilon) &=& -\alpha \left [ \ln \left (1+\frac{k}{\varepsilon}\right) + \ln \varepsilon \right ] \nonumber \\  
&\sim& -\alpha \left [\frac{k}{\varepsilon} + \ln \varepsilon \right]
\end{eqnarray*} 
and obtain 
\begin{eqnarray*}\label{e14}
p_k \sim \varepsilon^{- \alpha} \exp\left(- \alpha \frac{k}{\varepsilon} \right)
\end{eqnarray*}
Thus, \eqref{e7} is proportional to the exponential form $p_k \sim \exp(- \lambda k)$ with $\lambda = \frac{\alpha}{\alpha - 1}$.

\end{proof}

\begin{figure}[htp]
\begin{center}	
\includegraphics [width=8cm]{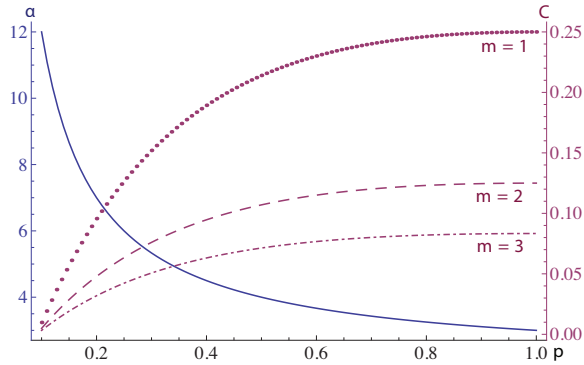}
\end{center}
\caption{Scaling exponent $\alpha$ for different values of $p$ (left frame); and global clustering coefficient $C$ for different values of $p$ and $m$ (right frame).} 
\label{fig:subfigures1}
\end{figure}

\noindent \emph{Remarks:} Theorem~$1$ implies that, as the network grows, the scaling exponent of the in-degree distribution depends on the stationary mean of forming triads. The distribution follows a strict power law for nodes with a degree greater than $\left( \alpha -1 \right) m$ and an exponential fit otherwise. The left frame of figure~\ref{fig:subfigures1} shows the value of the scaling exponent $\alpha$ for different values of $p$. Note that the mechanism generates network realizations with scaling exponent $\alpha \ge 3$. \\

\noindent \emph{Theorem $2$ (clustering coefficients): For all $\mathcal{G}_0$ that satisfy Assumption~$1$, the global clustering coefficient of $\mathcal{G}_t$ tends to~$C=\frac{p}{m (1+p)^2 }$ as $t \to \infty$. The asymptotic behavior of the local clustering coefficient for a node with in-degree $k_i = k$ follows }
\begin{eqnarray*}
C_i (k)=\frac{2\left(k + p m + \varepsilon  \ln \left (\frac{k + \varepsilon }{n + \varepsilon}  \right) (p-1) \right) }{\left(k+ p \varepsilon\right)\left(k+ p \varepsilon-1\right)}
\end{eqnarray*} 
\begin{proof}
Note that the only edge configuration to form transitive triplets is when node $j \notin \mathcal{H}_{t}$ attaches to $j' \in \mathcal{H}_{t}$ such that $j \in q_{j'}(t)$ and there exists a node $i \in \mathcal{H}_t$ such that $j' \in q_i (t)$. A triad is formed if node $j$ establishes a third edge to node $i$ that connects nodes $j$, $j'$, and $i$. The probability of establishing the third edge that closes the triplet is $p m$. Moreover, when node $j$ attached to the network, it connected (on average) to $m (1+p)$ outgoing neighbors (because node $j$ established $m$ edges according to the attachment process and then established an expected $p m$ additional edges according to the process of triad formation). Each outgoing neighbor of node $j$ also has (on average) $m (1+p)$ outgoing neighbors. Thus, there are $ m^2 (1+p)^2 $ different possible pairs to form triplets. The global clustering coefficient is given by 
\begin{eqnarray}\label{e9}
C=\frac{pm}{ m^2 (1+p)^2 }=\frac{p}{m (1+p)^2 }
\end{eqnarray}
which can also be expressed in terms of $\varepsilon$ as $C=\frac{\varepsilon - m}{\varepsilon^2}$.

\noindent Next, to capture the local clustering coefficient of a node $i$, note that the number of possible pairs of incoming and outgoing edges of node $i$ (with in-degree $k_i = k$) is given by
\begin{eqnarray}\label{e11}
&&{k + m(1+p) \choose 2} \nonumber \\
&=& \frac{(k+m(1+p))(k+m(1+p) -1))}{2}
\end{eqnarray}
Equation~\eqref{e11} captures the total number of possible triplets that involve node $i$. Now, to capture the number of actual triplets that involve node $i$, we consider three possible scenarios about the edges that may lead to triad formation: Node $i$ has $(i)$ two outgoing edges; $(ii)$ an outgoing edge and an incoming edge that was established through random attachment; and $(iii)$ two incoming edges with at least one of them having been established through triad formation. 

\noindent In scenario $(i)$, there are an expected
\begin{eqnarray}\label{e15}
p m
\end{eqnarray}
connected triplets.

\noindent In scenario $(ii)$, the number of incoming edges created through random attachment is
\begin{eqnarray}\label{e16}
\frac{dk^*_i (t)}{dt}=\frac{m}{N_0 + t -1}
\end{eqnarray}
with initial condition $k^*_i(t_i)=0$ (note that at $t=t_i$ the newly added node $i$ cannot have incoming edges that were established through random attachment). The solution to~\eqref{e16} is
\begin{eqnarray}\label{e17}
k^*_i (t)=m \ln \left( \frac{N_0 + t - 1}{N_0 + t_i - 1} \right)
\end{eqnarray}
Moreover, using~\eqref{e5} we also know that for node $i$ with in-degree $k_i (t)=k$
\begin{eqnarray}\label{e18}
\left( \frac{N_0 + t - 1}{N_0 + t_i - 1} \right)=\left( \frac{k + \left (1 + \frac{1}{p} \right) m}{n + \left (1 + \frac{1}{p} \right) m} \right)^{1+\frac{1}{p}}
\end{eqnarray}
Replacing~\eqref{e18} in~\eqref{e17} we know
\begin{eqnarray*}\label{e19}
k^*_i=\left (1 + \frac{1}{p} \right) m \ln \left( \frac{k + \left (1 + \frac{1}{p} \right) m}{n + \left (1 + \frac{1}{p} \right) m} \right)
\end{eqnarray*}
Note that the probability of establishing the third edge that closes the triplet is
\begin{eqnarray}\label{e20}
\left (1 + \frac{1}{p} \right) m \ln \left( \frac{k + \left (1 + \frac{1}{p} \right) m}{n + \left (1 + \frac{1}{p} \right) m} \right) p 
\end{eqnarray}
\noindent For scenario $(iii)$, the number of incoming edges that were established through triad formation is given by
\begin{eqnarray}\label{e21}
k- \left (1 + \frac{1}{p} \right) m \ln \left( \frac{k + \left (1 + \frac{1}{p} \right) m}{n + \left (1 + \frac{1}{p} \right) m} \right)  
\end{eqnarray}
which is the probability of establishing the third edge that closes the triplet. Finally, dividing the sum of~\eqref{e15}, \eqref{e20}, and \eqref{e21} by \eqref{e11}, we know  
\small{
\begin{eqnarray}\label{e22}
C_i(k) &=& \frac{2\left(k + p m \left (1+ \left (1 - \frac{1}{p^2} \right)  \ln \left (\frac{k + \left (1+\frac{1}{p}\right) m }{n + \left (1+\frac{1}{p} \right ) m}  \right) \right) \right)}{\left(k+ (1+p)m\right)\left(k+(1+p)m-1\right)} \nonumber \\
&=&\frac{2\left(k + p m + \varepsilon  \ln \left (\frac{k + \varepsilon }{n + \varepsilon}  \right) (p-1) \right) }{\left(k+ p \varepsilon\right)\left(k+ p \varepsilon-1\right)}
\end{eqnarray}}
where $\alpha=2+\frac{1}{p}$ and $\varepsilon=(\alpha - 1)m$.
\end{proof}

\noindent \emph{Remarks:} Theorem~$2$ implies that the values of $C$ and $C_i$ do neither depend on the initial network $\mathcal{G}_0$ nor the size of $\mathcal{G}_t$ (i.e., the clustering coefficients do not vanish as the network grows). The right frame of figure~\ref{fig:subfigures1} shows the value of the global clustering coefficient $C$ for different values of $p$ and $m$. Note that the model captures an inverse relationship between the clustering behavior and the amount of edges established during every attachment. 

\begin{figure}[htp]
\begin{center}	
\includegraphics [width=8cm]{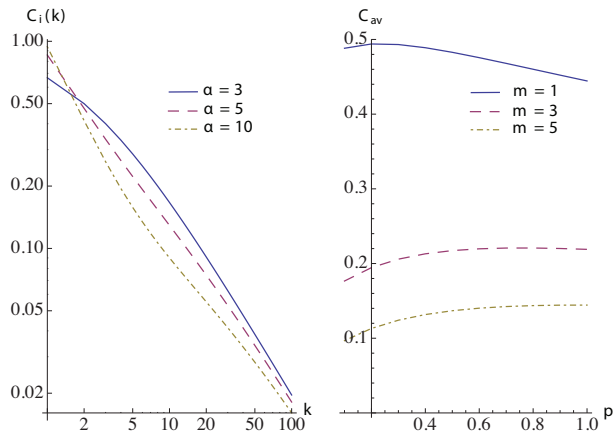}
\end{center}
\caption{Local clustering coefficient $C_i (k)$ for different values of $\alpha$ with $m = 1$ and $n=0$ (left); and average clustering coefficient $C_{av}$ for different values of $p$ and $m$ with $n=0$ (right).} 
\label{fig:subfigures2}
\end{figure}

The left plot of figure~\ref{fig:subfigures2} shows the effect of $\alpha$ on the local clustering coefficient. For nodes with a low degree, high values of $\alpha$ tend to form strong neighborhood clusters (below $p \varepsilon$). For nodes with a high degree, the effect is opposite and the local clustering coefficient is proportional to $k^{-1}$ (a behavior observed in empirical data \cite{Ravasz:2003fk}). Like for the global clustering coefficient, the right plot of figure~\ref{fig:subfigures2} shows an inverse relationship between the average clustering coefficient $C_{av}=\int_{n}^{\infty}p_kC_i(k)dk$ and the value of $m$. Finally, note also that the average clustering coefficient is slightly greater than the global clustering coefficient (also observed in empirical measures of clustering \cite{Newman:2003fv}).  

\section{Simulations}

To gain further insight into the network formation process, let~$N_0=12$ and $n=0$. Following similar ideas as in \cite{Jackson:2007uq}, \cite{Papoulis:1991fk}, let the probability of establishing additional links due to triad formation be $x_i (t) = 1-\frac{c}{u k_i (t)}$, where $u$ captures the compatibility between nodes and is chosen from a uniformly random distribution with support on $[0,1]$ (i.e., the random variable $X_t$ takes values $x_i (t)$). Let the parameter $c$, $0<c<u$, represent the cost of establishing additional links (here $c=0.1u$). The expected value of $X_t$ at time $t$ is given by
\begin{eqnarray}\label{e8}
p_t = E[X_t]=\int_{n}^{\infty}\int_{0}^{1}\left(1-\frac{c}{u k_i (t)} \right) p_u p_k  \ du \ dk_i
\end{eqnarray}
where $p_u = \frac{1}{u}$ and $p_k$ is the probability distribution of $k_i (t)$ according to Theorem~1. According to~\eqref{e8} it can be shown that because $p_t \to 1$ as $t \to \infty$, the process of triad formation has stationary mean $p=1$ for $n \ge 0$ and $m > 0$.

\begin{figure}[htp]
\begin{center}	
\includegraphics [width=8cm]{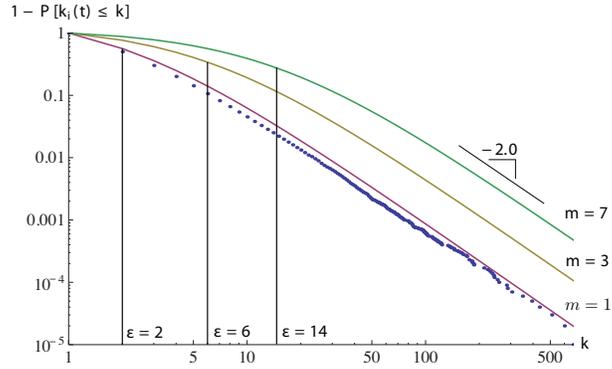} 
\end{center}
\caption{Complementary cumulative distribution function of the in-degree distribution $p_k$ on a logarithmic scale. The solid bottom curve represents the theoretical prediction according to~\eqref{e55} for $m = 1$; the dots represent simulation results. The two solid curves at the top represent predictions for values of $m = 7$ and $m = 3$.} 
\label{fig:subfigures3}
\end{figure}

Figure~\ref{fig:subfigures3} shows the in-degree distribution for different values of $m$ at $t=10^{5}$. For nodes with a low degree, the complementary cumulative degree distribution degenerates into the exponential form. In particular, the threshold $\varepsilon=2m$ characterizes the transition from an exponential (with $\lambda=\frac{3}{2}$) to a power law distribution (with $\alpha=3$).  

\begin{figure}[htp]
\begin{center}	
\includegraphics [width=8cm]{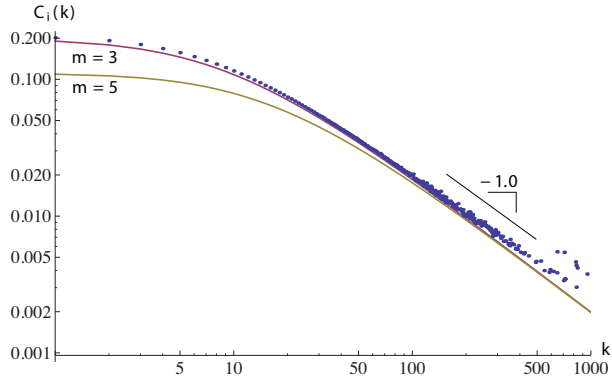}
\end{center}
\caption{Local clustering coefficient $C_i (k)$ as a function of the in-degree of a node. The solid top curve represents the theoretical prediction according to~\eqref{e22} for $m = 3$; the dots represent simulation results. The solid bottom curve represents the prediction for $m = 5$. } 
\label{fig:subfigures4}
\end{figure}

Figure~\ref{fig:subfigures4} shows the value of the local clustering coefficient $C_i(k)$ as a function of $k$ for different values of $m$. Note that the asymptotic expression of $C_i (k)$ tends to $k^{-1}$ for values greater than approximately $2m$. The fact that the degree of clustering that characterizes the different nodes follows a scaling law reveals the hierarchical organization of the generated networks \cite{Ravasz:2003fk}, \cite{Szabo:2003fk}. Note that the scaling of $C_i (k)$ emerges based solely on the process of triad formation (i.e., the tails of the distributions are the same for different values of $m$).   

\section{Group preference}

To explore the formation of communities, consider the following modifications to the node attachment and triad formation processes. Let the characterization of two types of nodes, denoted by $\delta \in \{ 1,2\}$, influence the formation of $\mathcal{G}_t$. The variable $\delta_i$ specifies the type of node $i$. 

\subsection{Node attachment}

Every time index $t$ a new node attaches to $m$ different nodes. The type $\delta_j$ of the new node $j \notin \mathcal{H}_{t-1}$ takes value 1 with probability $\frac{1}{2}$. When node $j$ attaches to the network, it connects to a node $j' \in \mathcal{H}_{t-1}$ of the \emph{same} type $(\delta_{j}=\delta_{j'})$ with probability $p_r$ (and with probability $1-p_r$ to a node of different type). 

\subsection{Triad formation}  

\begin{figure}[htp]
\begin{center}	
\includegraphics [width=8cm]{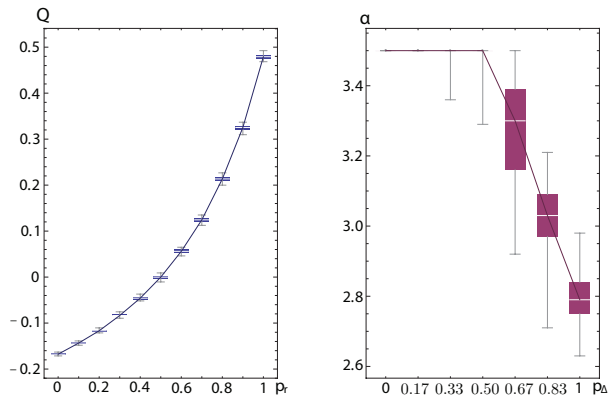}
\end{center}
\caption{Modularity $Q$ for different values of $p_r$ (left); and the resulting scaling exponent $\alpha$ for different values of $p_{\Delta}$ (right). } 
\label{fig:subfigures5}
\end{figure}

The probability $x_i(t)$ that node $j$ establishes an additional edge to an outgoing neighbor of node $j'$ (i.e., to some node $i$ such that $j' \in q_i (t)$) is also influenced by their type ($\delta_j$ and $\delta_i$). As before, $x_i(t)$ evolves according to a multivariate random variable $X_{t}^{\delta}$ with a finite expected probability $p_{t}^{\delta}$ and $\{X_{t}^{\delta}\}$ represents the random process associated with triad formation. Here, let
\begin{equation*}
x_{i}(t)=\left \lbrace 
\begin{array}{l}
	p_{\Delta} - \frac{c}{u k_i}, \quad \quad \: \; \mbox{if} \ \delta_j = \delta_i \\
     (1 - p_{\Delta}) - \frac{c}{u k_i}, \mbox{if} \ \delta_j \neq \delta_i \\
\end{array}
\right.
\end{equation*}
where $0 \le p_{\Delta} \le 1$. If nodes $j$ and $i$ are of the same type, the process of triad formation has stationary mean $p_{\Delta}$. Otherwise, it has stationary mean $1-p_{\Delta}$. The left plot of figure~\ref{fig:subfigures5} shows the modularity $Q$ for different values of $p_r$ for a network with~$N_0=12$,~$n=0$,~$m=1$,~$p_{\Delta}=0.5$ and $t=10^4$ (each point represents 100 simulation runs; error bars represent one standard deviation) \cite{Newman:2004fk}. The formation of non-overlapping communities is evident as $p_r$ increases. The right plot of figure~\ref{fig:subfigures5} illustrates the variation of the scaling exponent as $p_{\Delta}$ increases with $p_r = 1$. Note that when $p_{\Delta} \ge 0.5$ the scaling exponent starts to decrease, which indicates that group preference tends to heavily influence the power law behavior of the resulting network.  

\begin{figure}[ht!]
     \begin{center}
            \includegraphics[width=0.45\textwidth]{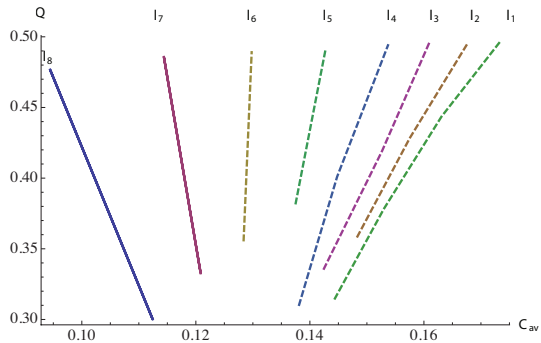}
    \end{center}
    \caption{Relationships between the modularity $Q$ and the average clustering coefficients $C_{av}$ for different values of $p_r$ and $p_{\Delta}$. Table~\ref{table1} shows the parameter regime for the various relationships.    
     }%
   \label{fig:subfigures6}
\end{figure}

We characterize the relationship between modularity $Q$ and the average clustering coefficient $C_{av}$ in figure~\ref{fig:subfigures6}. As indicated in Table~\ref{table1}, some parameter regimes produce linear relationships between $Q$ and $C_{av}$ (with different slopes for different values of $p_r$ and $p_{\Delta}$). For the relationships with a positive slope (i.e., $l_1 - l_5$) the model produces outcomes that resemble the empirical measures in \cite{Orman:2013fk}. 

\begin{table}[h!] 
\caption{Parameter regime for the correlations in figure~\ref{fig:subfigures6} when $n=0$.} 
\centering 
\begin{tabular}{c | c | c c c} 
\hline
& Slope &  Range of $p_r$ & & $p_{\Delta}$ \\
\hline 
$l_1$ & 6.268  &  [0.90,1]  & & 0.3  \\
\hline
$l_2$ & 7.060  &  [0.90,1] & & 0.4 \\
\hline
$l_3$ & 8.632  &  [0.90,1] & & 0.5 \\ 
\hline
$l_4$ & 11.707  &  [0.90,1]  & & 0.6 \\
\hline
$l_5$ &  20.773  &  [0.90,1] & & 0.7 \\
\hline
$l_6$ &  94.292  &   [0.80,1] & & 0.8\\
\hline
$l_7$ &  -23.726  &  [0.80,1] & & 0.9\\
\hline
$l_8$ &  -9.814  & [0.70,1] & & 1.0 \\
\hline
\end{tabular} 
\label{table1} 
\end{table}

\section{The U.S. Supreme Court citation network}

\begin{figure}[htp!]
\begin{center}	
\includegraphics [width=8cm]{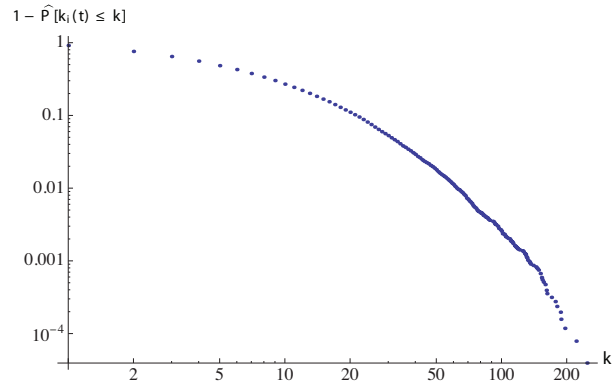}
\end{center}
\caption{Complementary cumulative probability for the in-degree distribution of the U.S. Supreme Court citation network; data from \cite{Fowler:2008fk}.} 
\label{fig:subfigures7}
\end{figure}

We apply the proposed mechanism to generate network realizations that resemble both the degree distribution and clustering properties of the U.S. Supreme Court citation network (using data from 1754 to 2002) \cite{Fowler:2008fk}. The citation network is created by a dynamic process, in which the number of opinions grows over time as judges write opinions that cite cases. The structure of the network captures which opinions get cited by later opinions. The evolution of the empirical network has the characteristic in-degree distribution shown in figure~{\ref{fig:subfigures7}}. It shows that the  Supreme Court opinion citations are concentrated in relatively small core of cases.

\begin{figure}[htp!]
\begin{center}	
\includegraphics [width=8cm]{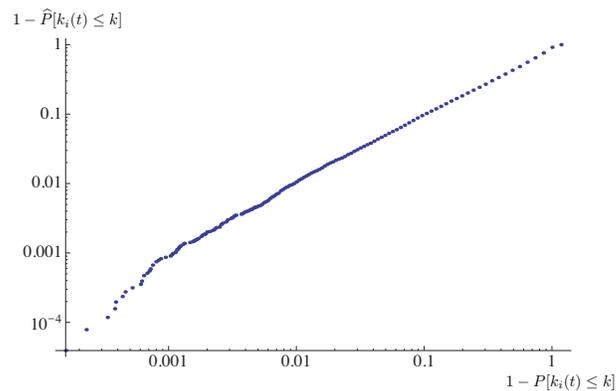}
\end{center}
\caption{Relationship between the empirical and the theoretical prediction of the degree distribution of the U.S. Supreme Court citation network.} 
\label{fig:subfigures8}
\end{figure}

Figure~{\ref{fig:subfigures8}} shows the relationship between the complementary cumulative in-degree distribution for the empirical network and its theoretical counterpart according to~\eqref{e55}. We use~$N_0=12$,~$n=0$,~$m=6$ and~$p=0.43$ which yields $\alpha=4.32$ and~$\varepsilon=19.92$. The empirical distribution from the data $1-\widehat{P}[k_i (t) \le k]$ correlates to the theoretical prediction $1-P[k_i (t) \le k]$ with a Pearson's correlation coefficient of $0.99$.  

Because the generated network does not rely on the dynamics of preferential attachment, it suggests that the formation of structure may be driven by the tendency of establishing additional citations and not a ``rich get richer'' dynamic \cite{Smith:2005fk}. Cases may accumulate legal authority (measured as the number of citations) not necessarily because -- having been cited approvingly be judges -- they are more likely to be cited in the future. This implies that if new citations are viewed as random, the characteristic structure will emerge, as long as each citation also refers to the opinions within the cases it cites.

\begin{figure}[htp!]
\begin{center}	
\includegraphics [width=8cm]{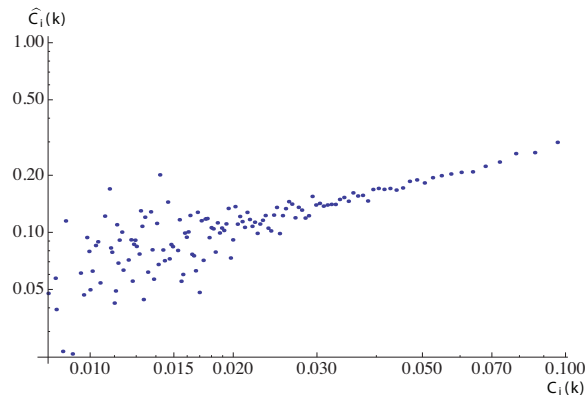}
\end{center}
\caption{Relationship between the empirical and the theoretical prediction of the local clustering coefficient of the U.S. Supreme Court citation network.} 
\label{fig:subfigures9}
\end{figure}

Next, figure~{\ref{fig:subfigures9}} illustrates the relationship between the empirical and the theoretical prediction according to~\eqref{e22}, of the U.S. Supreme Court citation network local clustering coefficient. The empirical measure from the data $\widehat{C}_i (k)$ correlates to the theoretical prediction $C_i (k)$ with a Pearson's correlation coefficient of $0.88$. 

When an opinion cites another one and the two opinions cite a third opinion (i.e., forms a triad that contributes to strong neighborhood clusters), it is a signal that these opinions (cases) are especially relevant to one another. These properties present an important source for discovering legal clusters that are tightly linked in terms of meaning and subject matter. 

\section{Conclusions} 
This paper introduces a mathematical framework that generates extended power law distributions with constant clustering coefficient based on a two-step mechanism: $(i)$ during attachment, a newly added node links to a finite number of randomly selected nodes; and $(ii)$ during triad formation, the new node may establish an additional link to one of the neighbors of the node it attaches to. The proposed mechanism is of interest because it helps explain the existence of extended power law networks with clustering properties that do not vanish as the size of the network grows. Generating network realizations with a desired scaling and clustering behavior allow us to evaluate which principles can lie behind the formation of relationships in large amounts of data. Moreover, our framework captures the effect of group preference in the formation of non-overlapping community structures. Analytical results about the processes that leads to overlapping community structures on clustered networks provides an important direction for future research.

\section*{Acknowledgments}
The authors thank the Office of Research at Pontificia Universidad Javeriana for the financial support under grant COFINPRO2011-3615.

\section*{References}
\bibliographystyle{unsrt}
\bibliography{Anteproyecto}

\end{document}